\documentclass[11pt]{article}
\usepackage[normalem]{ulem}
\usepackage{amsmath}
\newcommand{\stkout}[1]{\ifmmode\text{\sout{\ensuremath{#1}}}\else\sout{#1}\fi}

\usepackage{xcolor}

\usepackage{apacite}
\usepackage[sort&compress]{natbib} 
\usepackage{setspace}
\usepackage{graphicx}
\usepackage{wrapfig}
\usepackage{float}
\usepackage{multirow}
\usepackage[margin=0.75in]{geometry}

\title{Privacy preserving local analysis of digital trace data: A proof-of-concept}
\author{}
\begin{document}
\maketitle

Laura Boeschoten\footnote{\label{fn1}Dept. of Methodology and Statistics, Utrecht University, Utrecht, The Netherlands} \textsuperscript{*},
Adri\"enne Mendrik\footnote{\label{fn2}Eyra, Eyra Leap B.V., The Hague, The Netherlands},
Emiel van der Veen\textsuperscript{\ref{fn2}},
Jeroen Vloothuis\textsuperscript{\ref{fn2}},
Haili Hu \footnote{\label{fn3}Research Data Management, Utrecht University, Utrecht, The Netherlands},
Roos Voorvaart \textsuperscript{\ref{fn3}} and
Daniel L. Oberski \textsuperscript{\ref{fn1}} 

\bigskip

* Correspondence: l.boeschoten@uu.nl

\section*{Abstract}
We present PORT, a software platform for local data extraction and analysis of digital trace data. 
While digital trace data collected by private and public parties hold a huge potential for social-scientific discovery, their most useful parts have been unattainable for academic researchers due to privacy concerns and prohibitive API access. 
However, the EU General Data Protection Regulation (GDPR) grants all citizens the right to an electronic copy of their personal data. All major data controllers, such as social media platforms, banks, online shops, loyalty card systems and public transportation cards comply with this right by providing their clients with a  `Data Download Package' (DDP). Previously, a conceptual workflow was  introduced allowing citizens to donate their data to scientific- researchers. In this workflow, citizens' DDPs are processed locally on their machines before they are asked to provide informed consent to share a subset of the processed data with the researchers. In this paper, we present the newly developed  software PORT that implements the local processing part of this workflow, protecting privacy by shielding sensitive data from any contact with outside observers -- including the researchers themselves. Thus, PORT enables a host of potential applications of social data science to hitherto unobtainable data. 

\bigskip

\noindent \textbf{Keywords}

\noindent Data donation -- Digital trace data -- Privacy -- local processing -- data extraction -- software -- proof-of-concept

\section*{Introduction}
The General Data Protection Regulation (GDPR) grants all natural persons within the EU the right to an electronic copy of their personal data as collected by data controllers upon request. All major data controllers, such as social media platforms, banks, online shops, loyalty card systems and public transportation cards comply with this right by providing their clients with a `Data Download Package' (DDP). However, as only data subjects have the right to receive these DDPs, these digital traces can up until now not be accessed or used for scientific research. 

Recently, a workflow was introduced that allows researchers to analyse the digital traces found in these DDPs while preserving the privacy of the research participant \cite{boeschoten2020digital}. In this workflow, a research participant downloads the DDP of interest onto their personal device. Next, a \emph{local} processing step extracts only the features relevant for the research project from the DDP. After inspection and informed consent by the participant, these extracted features are then sent to the researcher who can then  perform the analysis of interest. 

In this paper, we introduce a proof of concept of the software that allows the local extraction step to take place in this workflow. More specifically, it is developed in such a way that researchers only need to share a URL with the participants. The local extraction step then takes place locally on the device of the participant within the web browser, and the extracted features are only shared with the researcher after consent is provided by the participant. In Table \ref{tab:1}, an overview of the main actors involved in the workflow can be found.

\begin{table}
  \begin{tabular}{p{4.5cm} p{7cm}} \hline
\textbf{General Data Protection Regulation (GDPR)}  & A legislation in the EU that (among other things) grants natural persons the right to receive their personal data from data controllers in a machine readable format\\ \hline
\textbf{Data Controller}                     & An entity, such as a tech company or public authority, that collects data on natural persons and is therefore required by the GDPR to provide that data in a DDP to the participant \\ \hline
\textbf{Data Download Package (DDP)}               & A machine readable electronic file  containing the personal data requested from the data controller by the participant\\ \hline
\textbf{Participant}                         & A person who is asked by the researcher to donate data from their DDP to the study \\ \hline
\textbf{Researcher}                          & A scientist who needs participant data encased in DDPs to answer a (social-scientific) research question \\ \hline
  \end{tabular}
  \caption{An overview and explanation of the key actors when digital trace data is donated for research   \label{tab:1}}
\end{table}

In the first section, we briefly explain the steps of the recently introduced workflow for data donation. Next, we describe how the software works broadly,  and discuss its technical specifications. Third, we illustrate the use of the software with two applications. Fourth, we discuss plans for future work and guidelines for ethical use, followed by a conclusion and discussion.

\section*{Methods}
In this section, we first describe step-by-step the procedure as perceived by the participant. We then describe in more detail the software that has been used to enable this procedure. We have named the software `PORT', as it serves as a `port' through which data is transferred from participants to researchers. We also describe the arrangements implemented to stimulate an open scientific approach by researchers when using PORT. 

\subsection*{Workflow}
\begin{figure}
\includegraphics[scale=0.265]{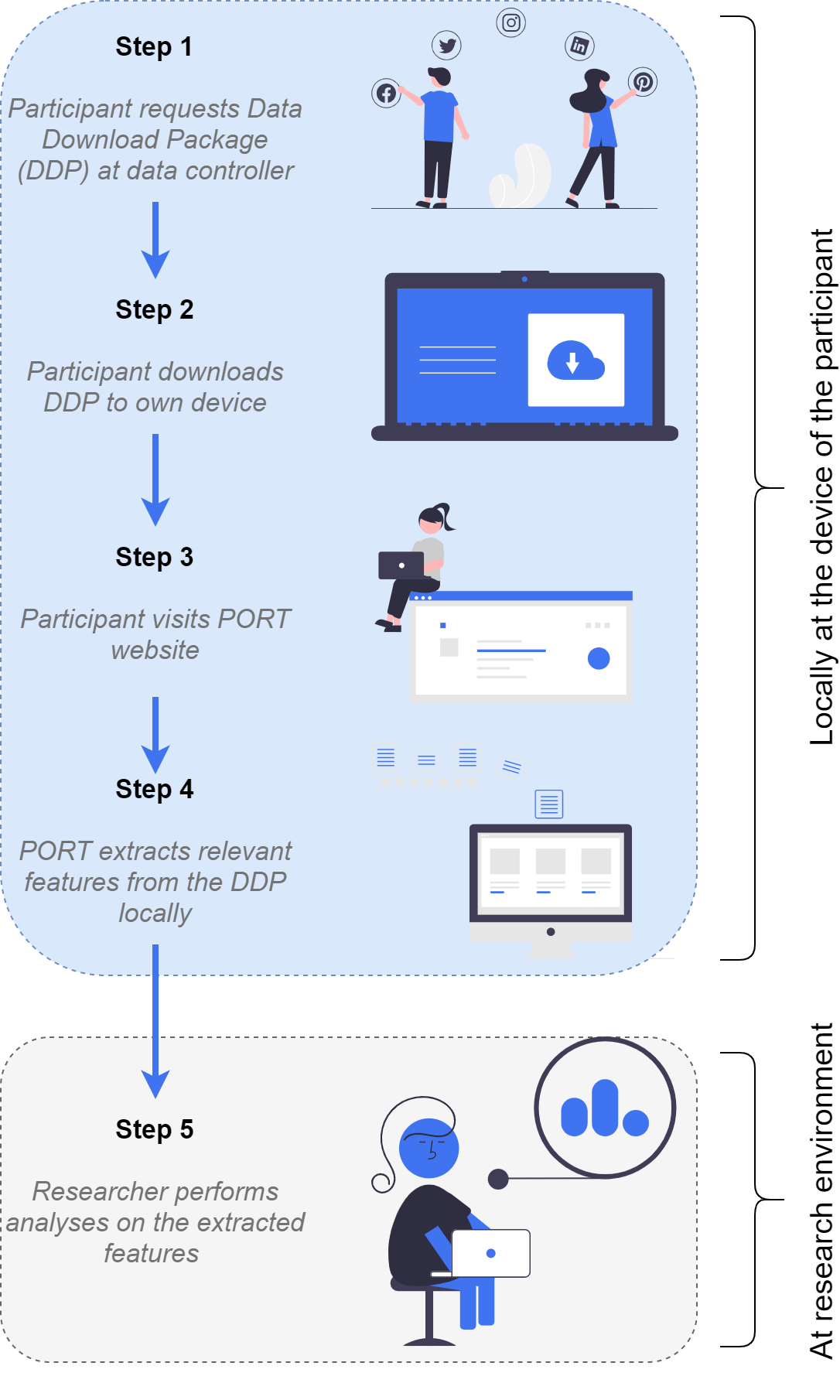}
\caption{{\bf Workflow.}
Step-by-step illustration of the workflow that allows for a privacy preserving analysis of Data  Download Package, locally at the device of a research participant.}
\label{fig:workflowDDP}
\end{figure}

When a participant is invited to participate in a research project that uses PORT to analyze digital trace data, the \textbf{first step} is always that the participant requests the DDP in which the researcher is interested from the respective data controller. It is the responsibility of the researcher to provide the participant with a clear, detailed and accurate description of how the DDP of interest can be requested, although many of the larger data controller companies already provide these descriptions on their websites\footnote{\url{https://www.facebook.com/help/1701730696756992}} \footnote{\url{https://faq.whatsapp.com/general/account-and-profile/how-to-request-your-account-information/}} \footnote{\url{https://help.twitter.com/nl/managing-your-account/how-to-download-your-twitter-archive}} \footnote{\url{https://support.snapchat.com/en-US/a/download-my-data}} \footnote{\url{https://help.instagram.com/181231772500920?helpref}}. The amount of time it takes for the participant to actually receive the requested DDP in practice varies between receiving it directly to taking a couple of weeks, depending on the size of the DDP, and the extent to which the data controller has automated the process. 

In the \textbf{second step}, the participant downloads the DDP to their own device. The way the participant receives the DDP can differ by data controller. In some cases, a file can be downloaded directly, and is stored locally on the device of the participant. At the time of writing, many banks, online stores, energy providers or public transport companies provide this option. However, in most cases, the preparation of the DDP by the data controller takes some time and the participants receive an e-mail or a notification within the platform of the data controller containing a download link. Google provides an option to place the latest version of the DDP on the participant's Google Drive and allows the user to update the DDP automatically.

After the DDP of interest is downloaded and stored on the device of the participant, the \textit{third step} is for the participant to visit the researchers' project website at PORT.

In \textbf{step four}, PORT extracts relevant information from the DDP; it does so locally and without communication with a remote server. In the next subsection, more detail regarding this procedure is provided. Note that up until this moment no information is shared with the researcher yet. Once PORT's extraction algorithm is finished, it it displays the extracted information and requests consent from the participant to share this with the researchers. If the participant consents, the extracted information is then encrypted and sent to the researcher for analysis. 

Once the researcher has obtained the extracted data of all participants, analyses can be performed to answer the research questions of interest in the \textbf{fifth step}. 

\subsection*{Software}
In the workflow, the extraction of relevant features of the digital trace data (the fourth step in Figure \ref{fig:workflowDDP}) is crucial. We ensured that this step takes place locally at the device of the participant, thereby preserving the privacy of the participant, by building PORT. PORT is a WebAssembly application that can run within browsers on both PCs and mobile devices.

\begin{figure}
\includegraphics[scale=0.265]{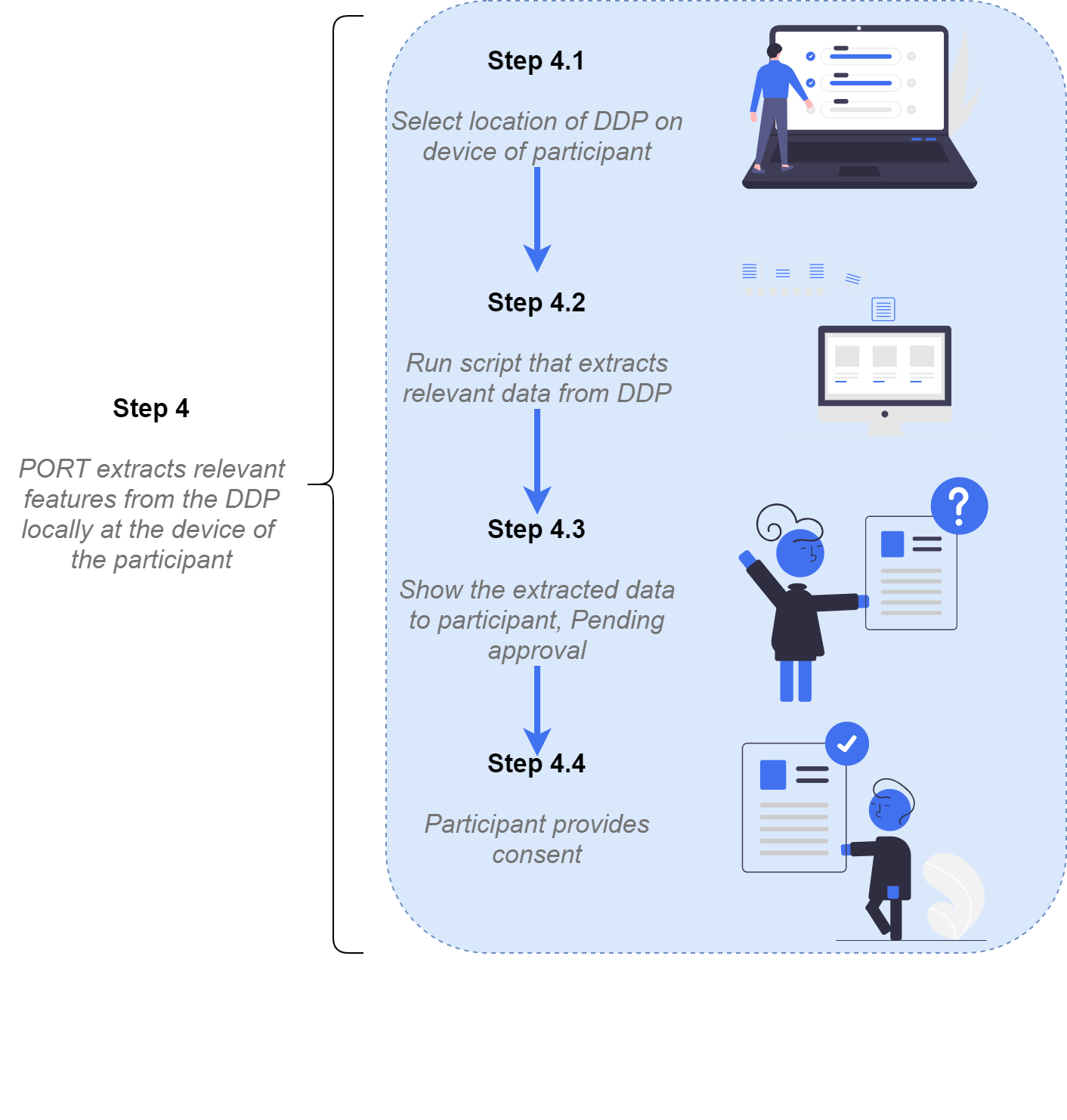}
\caption{{\bf Software.}
A more detailed illustration of the steps taken by a research participant during the local processing phase of the proposed workflow, which can be performed using the PORT software.}
\label{fig:workflowDDPsoftware}
\end{figure}

Figure \ref{fig:workflowDDPsoftware} provides more detail regarding  PORT's functionality.  The Figure shows that, first (\textbf{step 4.1}), the participant selects the location in which the DDP is stored at the device. 

Once the DDP is selected, PORT runs a script in the browser that locally extracts the data from the DDP that is relevant for the researchers (\textbf{step 4.2}). The data extraction script consists of code written by the researcher in Python and it is tailored to the specific research question under investigation, and to the specific DDP that is used to answer this question. In the next section, this idea is illustrated using several scripts and different DDPs. 

PORT makes use of Pyodide \cite{Pyodide}, an open-source library that enables running Python in a web browser through having the Python interpreter compiled to WebAssembly \cite{WebAssembly}. WebAssembly is an open standard for portable binary-code format, enabling high-performance applications on the web. Thus, the custom Python scripts will run in a safe, sandboxed environment, while the browser's security and permission policies are enforced. When the participant clicks the ``process" button in PORT, the DDP is sent to the WebWorker. Then the JavaScript code presents the DDP through the Pyodide bridge API as a regular Python file object. A function in the Python code is invoked which receives this file object and interacts with it.

Once the function has extracted the relevant data it returns the extracted data accompanied by information about what exactly has been extracted. This data is then converted by the Pyodide bridge to JavaScript object types. The JavaScript within the WebWorker creates an event in Java and notifies the primary JavaScript. This JavaScript code then updates the browser domain \cite{DocumentObjectModel} to present the extracted data to the participant (\textbf{step 4.3}). Conceptually, this step can be considered as if a small system environment is built within the browser, completely separated from the device on which it is run. This is a safe procedure as the environment is able to process the data that is uploaded in the environment, but is not able to access anything other present on the device. In addition, this system environment is destroyed as soon as the browser page closes. 

In the final step (\textbf{step 4.4}), the participant  inspects the extracted data and can provide informed consent for sharing this data with the researchers by clicking the respective button. If the participant has freely chosen to do so, the extracted data is shared with the researchers over the internet; note that this is the first communication of the participant's device with a server controlled by the researchers. Alternatively, this moment can also be used to decline consent. PORT currently uses an Elixir, Phoenix server hosted by Eyra B.V.. For future use by applied researchers, integration with preferred online repositories or servers hosted by academic institutions is also possible, allowing for the extracted data to be send directly to the location of interest. If this is possible and how the connection looks like in practice, depends on the institution receiving the data. We provide guidelines and best practices in the subsequent section.

\section*{Example applications}
In this section, we illustrate how PORT can be used to extract features from DDPs to collect digital trace data for answering research questions. We provide two applications, in which we select a suitable DDP and develop a Python script that extracts the information needed from the DDP. We initially developed the script by inspecting  our personal DDPs. Based on the data structures found here, we simulated a DDP. The simulated DDP is then used to illustrate how PORT is able to locally extract features from it. Next, PORT shows the extracted features, after which informed consent can be provided. After informed consent, the extracted features are send to the repository of the study, currently at the server of Eyra B.V.. The simulated DDPs, the code used to simulate the DDPs and the Python scripts integrated in PORT for these studies can be found at \url{https://github.com/UtrechtUniversity/port-poc}. Because we used simulated DDPs, the steps in the applications are fully reproducible.

\subsection*{Measuring differences in where time is spent during Covid-19 Lockdowns}
The first application focuses on the hypothetical research question, ``how does travel behavior change in times of a Covid-19 lockdown?". This research question could be of interest to policy makers or academic researchers. Once supplemented with participant's background characteristics, the same data could also be used to investigate how individual characteristics related to movement behavior during the pandemic. 

Answering this research question requires location data and the amount of time the person spends at the respective locations. Since self-reporting instruments such as time-use surveys or diaries are prone to recall bias \cite{elevelt2021you}, digital trace data are an interesting alternative. DDPs are particularly relevant here, as the researcher is interested in retrospective data. Such past data are difficult to obtain using, for example, wearables:  the researcher has to anticipate that `something interesting' will happen in the future. In contrast, DDPs that provide location data generally do so for a considerable portion of the past.

When exploring different data controllers that store location information, we focus on DDPs collected by the Android operating system, as Android has the majority market share \cite{Statscounter}, which means that it is more likely that a participant actually has the DDP. In more extensive studies, researchers can consider to also collect DDPs that store location information via for example iOS, to reduce the amount of missing data that is induced due to the participants not having an Android DDP. In the variety of DDPs collected by Android and Google (which can be found under `Google Takeout'), the so-called `Google Semantic Location History' (GSLH) contains monthly .JSON files with information on geolocations, addresses, time spend in places, activity and more. 

\subsubsection*{Google Semantic Location History data extraction} 
Information on time spent at locations and travelled distances is easily extracted from the Google Semantic Location History DDP, as it contains start and end time of each visited place and activity, and the travelled distance per activity. We simply need to select and sum the appropriate values per month. Note that the DDP contains time information in milliseconds since January 1, 1970, and distance information in meters. For readability, we convert these to days and kilometers, respectively.

Since we generated a synthetic version of the Google Semantic Location History DDP (see Appendix \ref{appendix:A}), we can check if the extracted information is consistent with what we specified in our data faker. Figure \ref{fig:dd3} shows that we generated 49, 47 and 19 visited places for 2019, 2020, and 2021, respectively. These numbers are close, but not equal to, the number of addresses that were given as input (50, 50, 20). This is because we selected the visited places with weighted probabilities from the input list. Furthermore, the amount of time spend in places is 80\% in 2019 and 2020, and 95\% in 2021 as expected.

\subsubsection*{Donating Google Semantic Location History data using PORT}
\begin{figure}
\includegraphics[scale=0.85]{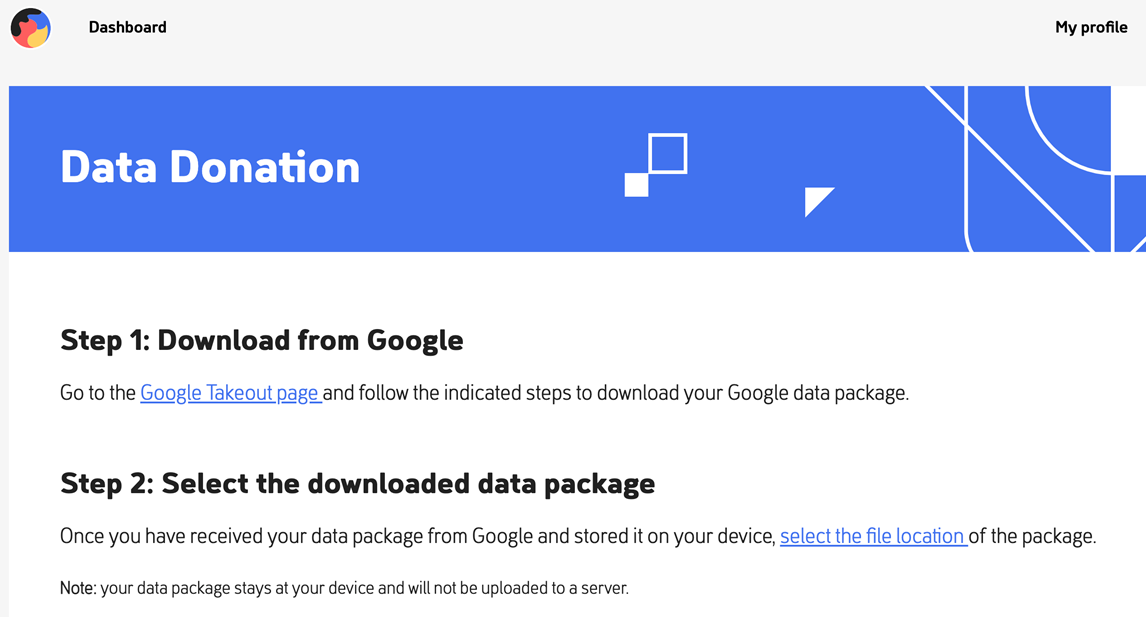}
\caption{Once the participant has downloaded their GSLH or other DDP, they can select the file location on their device for data donation using PORT.}
\label{fig:dd1}
\end{figure}

When participants are invited to participate in this study, the first step is that they are requested to download their personal DDP, as can be seen in Figure \ref{fig:dd1}. First, the participant goes to their Google profile, for example by going to \url{https://www.google.com} and then click on the profile button at the right upper corner. Here, they click the button `Manage your Google Account'. Once they arrive at their Google Account, they can click the tab `Data and personalization', where they can then click on the tile `manage your data and personalization'. In this new window, they can click the button `Download your data'. This takes the participant to the `Google Takeout' section. Here, a list is shown containing all Google products for which a DDP can be downloaded. Here, it is most convenient to first click on `deselect all' and then re-select only the `Location History'. This by default contains both Google Location History and Google Semantic Location History in .JSON format, although it is also possible to deselect one of these DDPs or to change the format into .Kml. Next, the participant clicks the button `Next step' and then `Create export'. Once Google is finished with preparing the DDP, the participant receives an email guiding them towards the same page that now contains a button `Download'. If the participant clicks this button, the DDP will be stored at their device. Note that this is the sequence of steps at the time of writing and can be subject to changes over time. 

\begin{figure}
\includegraphics[scale=0.60]{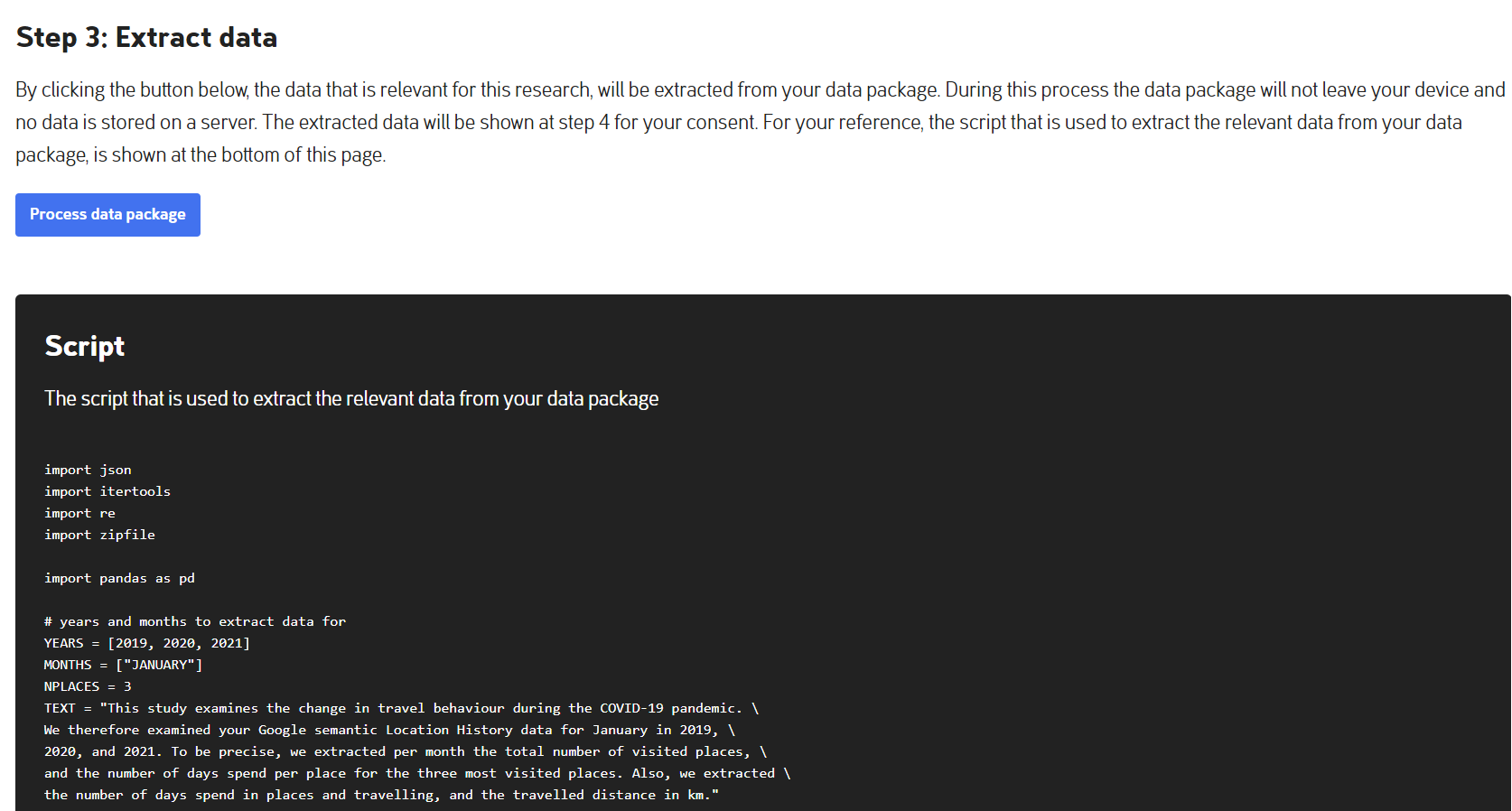}
\caption{With the `process data package' button in PORT, the GSLH Python extraction script extracts the relevant features from the selected DDP. For transparency, PORT shows the complete Python script.}
\label{fig:dd2}
\end{figure}

Once the Google Semantic Location History DDP is downloaded and stored, the participant can click on the `select the file location' button at PORT as seen in Figure \ref{fig:dd1}. Once the DDP is uploaded in PORT, the third step becomes visible on screen, as seen in Figure \ref{fig:dd2}. At this step, for transparency the extraction script is visible for inspection. Once the participant clicks the `Process data package' button, the extraction script is run on the selected DDP. 

\begin{figure}
\includegraphics[scale=0.55]{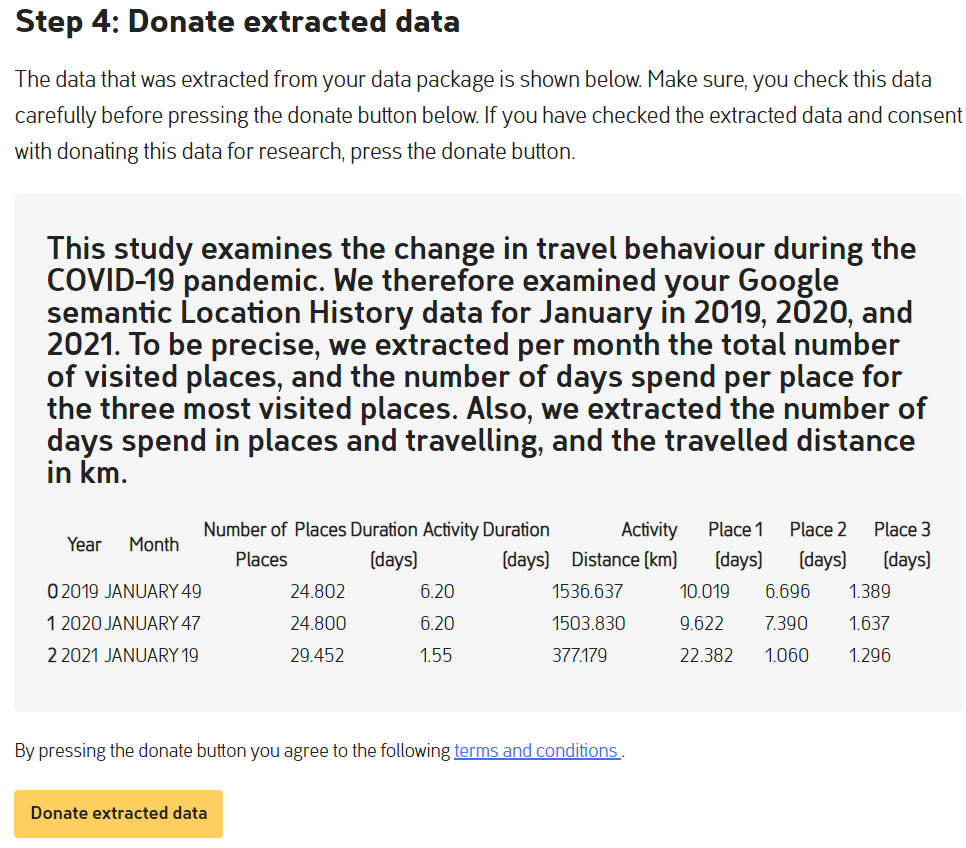}
\caption{The extracted data from the Google Semantic Location History DDP as displayed to the participant. Below the extracted data, the participant can find the `Donate extracted data' button.}
\label{fig:dd3}
\end{figure}

Once the script is finished, the extracted information is shown to the participant including an explanation of what the various printed numbers exactly mean, as can be seen in Figure \ref{fig:dd3}. Note that the extracted data here contains no personal information. The addresses of the top three visited places are replaced with identifiers Place 1, Place 2, and Place 3. If different addresses are in the top three in different years, place identifiers with increasing numbers will be used. In other words, the place identifiers are anonymous replacements of the addresses, and we only show the top three visited places of each year.

Below the extracted information, there is a button `Donate extracted data'. When the participant clicks this button, the extracted data is transmitted to the server of the researcher to which PORT is connected and the complete DDP was never shared with the researchers.

\subsection*{Measuring differences in online news consumption due to the Covid-19 induced curfew}
The second application focuses on the research question ``How does news consumption changes during times of a Covid-19 curfew?", which is closely related to the research conducted by \cite{broersma2021novel} on changes of news habits during the Covid-19 pandemic. Data about a person's page visits to news websites in relation to this person's general browser behavior can help to answer this research question. Google Chrome is the most frequently used browser with a 65\% market share \cite{Statscounter2}. Therefore, it makes sense to again obtain the information that can help to answer this research question via the DDP of Google, Google Takeout. Within this DDP, the browsing history of the browser Google Chrome is found in the file `BrowserHistory.JSON'. This .JSON file contains a user's entire compiled browser history, from the moment they started using Google Chrome or since the last time they deleted their browsing history. 

\subsubsection*{Google Search History data extraction}
Similar to the previous application, the DDP was simulated. More detail can be found in Appendix \ref{appendix:B}. Analyzing the search behavior as listed in the BrowserHistory.JSON file consists of a number of steps. First, the ``BrowserHistory.JSON" file is extracted from the provided Google ``Takeout.zip"-file, and is loaded as dictionary. Using the timestamp provided for every search in the ``BrowserHistory.JSON" file (\emph{time\_usec}), it can be determined in what period the website was visited: i.e, before the start of the curfew, during the curfew, or after the curfew. 

\begin{figure}
\includegraphics[scale=0.30]{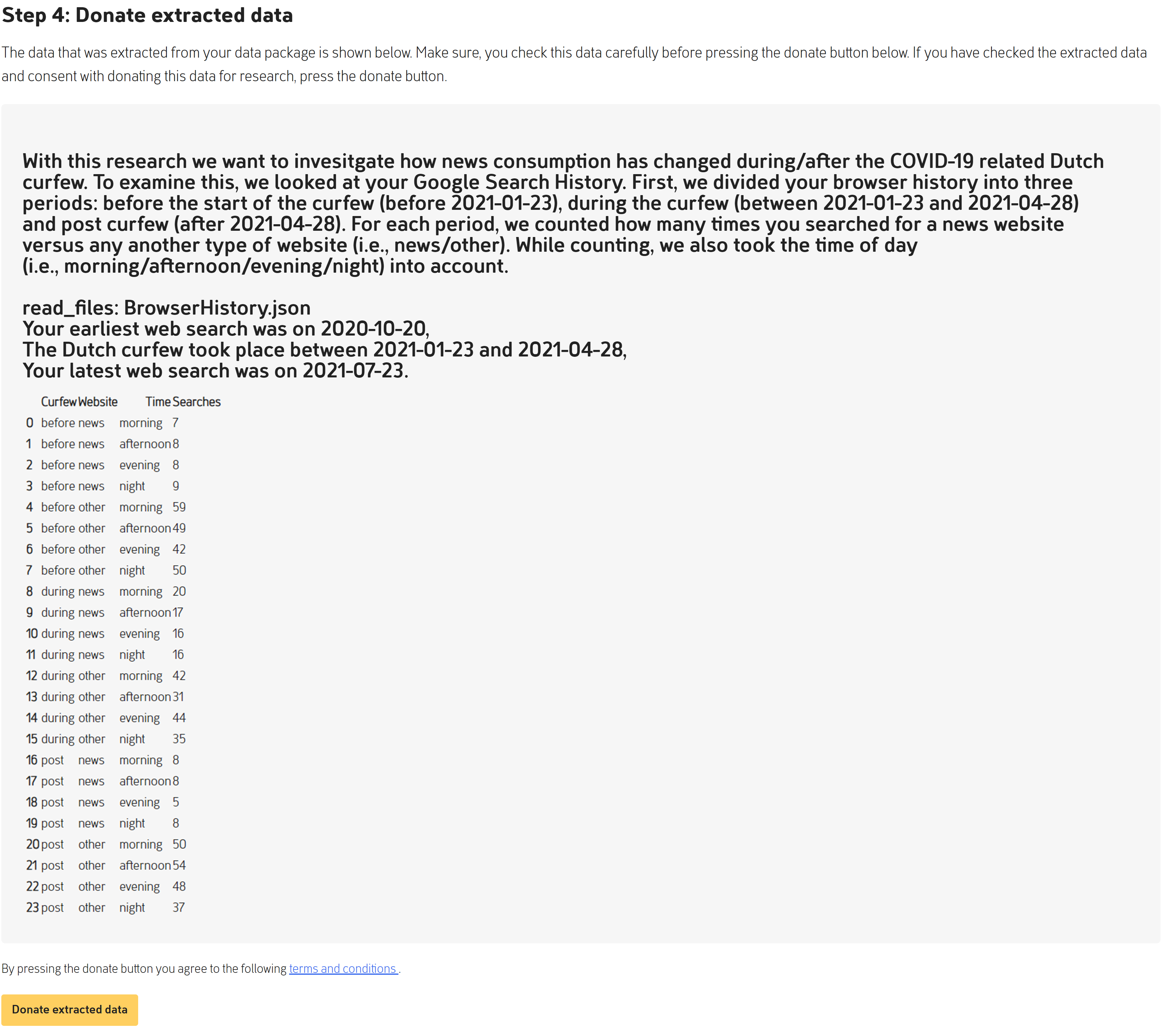}
\caption{The extracted data from the Browser History DDP as displayed to the participant. Below the extracted data, the participant can find the `Donate extracted data' button.}
\label{fig:dd5}
\end{figure}

Second, it is determined whether a certain website is a news-site or not, and for both groups the number of individual visits are counted. To determine whether a website is a news-site or not, each website visit's \emph{url} is matched with a list of the most popular Dutch news websites according to Wikipedia \cite{Wikipedia2021Nieuws}. In addition, for each website visit, it is determined whether the visit took place before, during or after curfew, and during morning, afternoon, evening or night, based on the supplemented timestamp. Finally, all information is combined into a single table, where each row represents a different profile (e.g. newssites/before curfew/morning) and sums the total number of individual website visits corresponding to that profile (see also Figure \ref{fig:dd5}. When this information is collected from a group of participant, a researcher can determine whether news-consumption increases during a period of curfew, especially in the evening and night, compared to periods without a curfew.

\subsubsection*{Donating Google Search History data using PORT}

Similar to the first application, participants start by requesting their personal DDP via their Google profile, as we could see in Figure \ref{fig:dd1}. However, instead of requesting their `Google Location History', they now request their Chrome `Browser History'. Again, the participant clicks on `Create export' and receives an email when the file is ready for download. Once the participant has downloaded the `BrowserHistory.JSON' file and stored them at their device, they can continue to the PORT website. 

\begin{figure}
\includegraphics[scale=0.35]{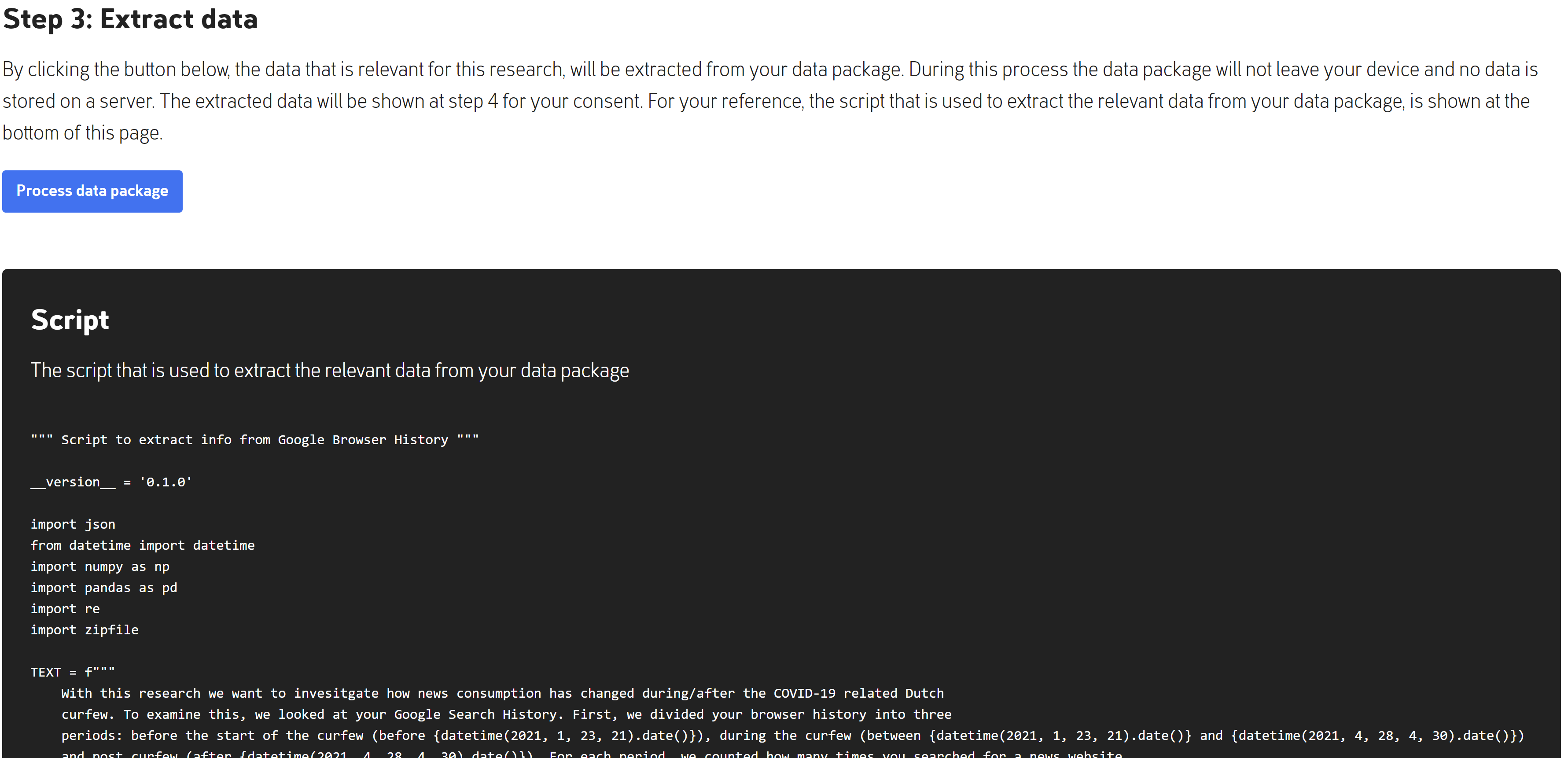}
\caption{With the `process data package' button in PORT, the Browser History Python extraction script extracts the relevant features from the selected DDP. For transparency, PORT shows the complete Python script.}
\label{fig:dd4}
\end{figure}

In PORT, the participant then again can click the `select the file location' button (see Figure \ref{fig:dd1}). Once the DDP is uploaded, the third step becomes visible on screen. As can be seen in Figure \ref{fig:dd4}. Here, the Python extraction script that has been written for this specific application is now shown, as such that it can be again inspected by the participant if this is of interest. Once the participant clicks the `Process data package' button, this extraction script is run on the selected DDP. 

The extraction information that is shown to the participant can be seen in \ref{fig:dd3}. Again, a detailed description of what can be found in the data is included. From the explanation and extracted data it can be seen that the extracted data contains no personal information. The output then describes the earliest web search found within the DDP, the period of the curfew, and the date of the latest web search. Next, a frequency table is shown. This table describes how many web visits were made per period (i.e., before, during, or after curfew), per website type (i.e., news or other), per time of day (i.e., morning, afternoon, evening, or night). Once the participant clicks the button `Donate extracted data', only this table is transmitted to the server to which PORT is connected that can be accessed by the researcher and the complete DDP was never shared with the researchers.

\section*{Future work and guidelines for ethical use}
Although we are currently only presenting a proof-of-concept of how PORT operates, we want to provide a first set of guidelines to help researchers use PORT in a legal and ethical way. To guide you through our considerations, we discuss the process of using PORT as a researcher in three phases: (1) the participant recruitment phase; (2) the local processing phase and (3) the data extraction phase.

The first contact between the participant and the researcher is made at the \textbf{participant recruitment phase}. To enable recruitment of participant, a researcher might use a professional participant recruitment platform or collaborate with an existing panel study. Regardless of the exact structure, it is important to keep in mind that the researcher is responsible for generating unique identifying keys for each participant, that can be used in such a way that once the extracted features are stored, they are labelled with this key and can for example be used for linkage to survey measures, or to make conclusions regarding the selectivity of the final obtained sample. At this stage, it is good practice to inform participants regarding the study aims, the type of data they will exactly share and the process they will go through similar to any other study. 

For the \textbf{local processing phase}, the researcher prepares a Python script. This Python script should adhere to a number of rules. First, it should extract only the features from the DDP that are relevant for the particular research question. Note that if these extracted features are sensitive, the collected data should be treated as such in the subsequent steps, e.g. the safety measures taken at the data storage facility can be related to this. Second, the script should be able to handle variability in structure and content over DDPs from different research participants. Third, the script should present the extracted features in a clear and intuitive way for the participant to review when providing informed consent, and for the researcher to process for further analysis, including linkage to the data from other participants or linkage of other data sources regarding the same participant, for example survey data. Regarding the presentation of extracted features to the participant, the researcher can present the features in an intuitive way, using a table, a graph or a figure such as a map. However, all features shared with the researcher should have been presented to the participant. 

When using PORT, we encourage researchers to adhere to the FAIR principles \cite{wilkinson2016fair} while retaining the preservation of privacy if the type of features extracted demands this. To encourage \textit{findability} of the research projects using PORT, a separate project page is available on the website of PORT for each project, which can be accessed from the starting page. At the project page, information about researchers and research findings can be shared, which can be updated throughout the duration of the project. Furthermore, the Python script used to extract features is made publicly available here, so that it can be re-used for other projects. At last, information regarding the location where the data will be stored should be provided. 

\textit{Accessibility} of the extracted features very much depends on the sensitivity of its content and different options are possible. Researchers can for example choose to let the collected data be collected at the Eyra server, or to let the extracted features be send directly to a repository in a protected cloud environment. If differs per repository what is exactly possible. 

As the nature of PORT is \textit{interoperable}, it can in theory be combined with different participant recruitment and data storage platforms. However, whether a particular platform of interest by the researcher can be combined in practice depends on platform specific characteristics. 

The prototype of the software used for PORT is developed open sourced (see \url{https://github.com/eyra/port-poc}) and we ask users to share their developed Python scripts open source as well such that they can be \textit{reused}. Furthermore, once researchers inspected and cleaned the collected data where needed, they can provide information for other researchers if and how their collected data can be \textit{reused} for other studies. 

\section*{Conclusion and Discussion}

In this paper, a proof-of-concept is introduced of software that locally, on the device of a research participant, extracts relevant features from Data Download Packages (DDPs). Thanks to this software, it is now possible to use digital trace data for research purposes with data that are not publicly available, in an ethical manner. The software allows for privacy preservation of research participants and for true informed consent. 

To use the software for research, it should always be supplemented with a method to recruit participants and with a location where the extracted data can be stored. Currently, the software is hosted, which means that the university of the respective researcher should have a processing agreement with hosting company, Eyra. In future, integration of the software with existing panel studies such as the LISS panel \cite{scherpenzeel2018true} or Understanding Society \cite{buck2011understanding} may be of interest. In addition, it is then probably desired that the extracted data is directly sent to the server of the respective organization. Alternatively, researchers might be interested in storing the extracted data directly in a repository for research data, hosted by organizations such as DataVerse \cite{king2007introduction} or Surfdrive \cite{Surf}. 

When using the workflow as proposed by \cite{boeschoten2020digital} and the software as presented in this paper for a research project, the only task that remains is that data scientists and applied researchers collaborate to develop a high-quality extraction script in Python that is flexible in terms of handling a variety of data structures. See \cite{boeschoten2021automatic} for an overview of the data types and structures that are typically found in DDPs from major data controllers. When developing such an extraction script, it is important to find a balance between ensuring on the one hand that all information relevant for answering the research question of interest is extracted, while on the other hand no sensitive sensitive data is unnecessarily collected.

In addition, it should be noted that using the proposed workflow and additional software is not suitable for every type of research question. For example, in a research project of a more exploratory nature where the researcher aims to find out what type of data can be found in DDPs, or when a researcher aims to develop and evaluate the performance of an extraction or prediction algorithm, an extraction script might not be suitable. However, even in such situations a researcher can consider to use the workflow and software, but instead of using an extraction script, a de-identification script can be more appropriate. An example of such a de-identification script has been developed by \cite{boeschoten2021automatic} for Instagram DDPs, which only selects the files within the DDP that are of interest for the research and then removes all identifiers from these files. When applying the workflow in such a way, two main study principles are still applied: the privacy of research participants is protected and only the necessary data is collected. 

In future research, we plan to  integrate our system with existing participant recruitment platforms and data archives. In addition, we of course plan to apply the workflow and software for practical research questions and develop more extraction scripts. Furthermore, we hope to focus on a more intuitive way of presenting the extracted data to participants, as such that the extracted information is insightful and useful for the participant, which hopefully leads to more engaged participants. Other expansions of interest are extending the number of Python packages that can be used in the extraction script, including options for visual content, and also a possibility to develop extraction scripts using the \texttt{R} programming language.

\medskip
To conclude, PORT opens up vast new research opportunities for researchers with an interest in digital trace data. Digital trace data can be collected in DDPs by a substantive part of the world's population and regarding every aspect of their (digital) lives, such as social media, banks, online shops and shops with loyalty card systems, travel and movement behavior, and health. PORT allows for a privacy-preserving analysis of these  digital traces for research purposes, subject to  true informed consent.

\bibliographystyle{apacite}
\bibliography{references}

\appendix

\section{Google Semantic Location History data simulation}\label{appendix:A}
The code used to generate the simulated GSLH DDP builds upon the open source Python packages ``GenSON" \cite{GenSON2021}, ``Faker" \cite{Faker2021}, and ``Faker-Schema" \cite{FakerSchema2021}. GenSON is used to generate a .JSON schema of a personal DDP. The .JSON schema describes the format of the GSLH data, and can be used to generate dummy data with a similar format. Dummy data is then generated by converting the .JSON schema to a custom schema expected by faker-schema in the form of a dictionary, where the keys are field names and the values are the types of the fields. The values represent available data types in Faker, packed in so-called providers. Faker provides a wide variety of data types via providers, for example for names, addresses, and geographical data. This allows us to easily customize the dummy data to our specifications.

We generated GSLH dummy data for the years 2019, 2020, and 2021. First we created 50 dummy addresses with Faker, with coordinates in the Netherlands within a radius of 0.1 degrees of each other. For the years 2019 and 2020, the visited locations were randomly chosen with weighted probabilities from all 50 addresses, for 2021 only 20 of these addresses were available for random selection. The weighted probabilities were used to create a top three of most frequently visited locations. We also varied the duration of time spent at locations per year, where in 2019 and 2020 80\% of the time is spent at locations and in 2021 95\% of the time is spent at locations. The geodesic distances between visited locations were calculated in kilometer using ``GeoPy" \cite{GeoPy2021}.

\section{Google Search History data simulation}\label{appendix:B}
The BrowserHistory.JSON file is structured in such a way that, for each web search, information about that specific web search is listed in a dictionary. Such a dictionary entails information about how the website was visited (\emph{page\_transition}, i.e., `LINK', `GENERATED', or `RELOAD'), the title of the web page (\emph{title}), the complete url to the weblink (e.g. \url{https://nos.nl/}), a unique identifier for the user (\emph{client\_id}), and the timestamp of the visit (\emph{time\_usec}). As the BrowserHistory.JSON file is a mere compilation of all these unique search dictionaries stored within one overarching `Browser History' dictionary, the code used to generate the simulated data is quite straight forward.

First, an empty dictionary is created with the above mentioned structure. For each simulated web visit, a random \emph{page\_transition} (i.e., `LINK', `GENERATED', or `RELOAD') is chosen using the ``Random" package in Python \cite{Random2021}, which uses the Mersenne Twister algorithm \cite{matsumoto1998mersenne}. Second, the \emph{title} and \emph{url} of the web page are constructed using the sentence() and website() functions of the open source Python packages ``faker" \cite{Faker2021} and ``random" \cite{Random2021}, respectively. After the URLs are created, the homepage part of the URL is replaced with a randomly selected Dutch news website for n\% of the URL. These news sites are randomly picked from a sampled list of popular Dutch news website as obtained from Wikipedia \cite{Wikipedia2021Nieuws}. The \emph{client\_id} is created using a random set of letters and digits, making up a string of length 10. Finally, timestamps are generated in such a way that morning, afternoon, evening, and night times occur evenly. 

Note that the simulated browser history data is created in such a way that there is a difference in total number of news versus random website visits before (before 2021-01-23), during (between 2021-01-23 and 2021-04-28), and after (after 2021-04-28) the curfew. The default settings will result in a `Takeout.zip' file, with a sub-folder 'Chrome' leading to a `BrowserHistory.JSON' file, i.e., a Browser History dictionary containing a 1000 web searches. During the period of the curfew - compared to the period before or after the curfew - there were 15\% more news web visits and, overall, more web visits during the evening.

\end{document}